\documentclass[english]{article}
\usepackage[latin9]{inputenc}
\usepackage{geometry}
\usepackage{amsmath}
\usepackage{amssymb}
\usepackage{graphicx}
\usepackage{babel}
\usepackage{placeins}
\usepackage{caption}
\usepackage{fullpage}

\begin{document}

\title{Matrix Model of Strength Distribution: Extension and Phase Transition}

\author{Arun Kingan and Larry Zamick \\
 Department of Physics and Astronomy\\
 Rutgers University, Piscataway, New Jersey 08854 }
\maketitle
\begin{abstract}
In this work we extend a previous study of matrix models of strength
distributions. We still retain the nearest neighbor coupling mode
but we extend the values the coupling parameter v. We consider extremes,
from very small v to very large v. We first use the same transition
operator as before \textless{n T(n+1)\textgreater{}} = constant (=1). For
this case we get an exponential decrease for small v, as expected, but we
get a phase transition beyond v=10, where we get separate exponentials
for even n and for odd n. We now also consider the dipole choice where
\textless{}nT(n+1)\textgreater{} = $\sqrt{(n+1)}$.
\\
\\
 PACs Number: 21.60.Cs 
\end{abstract}

\section{Introduction}

This work is a continuation of work done before on matrix models of
strength distributions {[}1{]}. Matrix mechanics was of course introduced
into quantum mechanics by Heisenberg{[}2{]} and Born and Jordan{[}3{]}.
In the previous work we had a matrix in which the diagonal elements
were E$_{n}$  = nE with E=1. We introduced a constant coupling v which
for a level E$_{n}$ occurs only with the nearest neighbors E$_{(n-1)}$
and E$_{(n+1)}$. Note that the only relevant parameter is v/E.

The matrix Hamiltonian that we will use is the same as that used in ref[1]. It is shown
here again:

\setcounter{MaxMatrixCols}{15}

\begin{gather*}
H=
\begin{bmatrix}
0 & v & 0 & 0 & 0 & 0 & 0 & 0 & 0 & 0 & 0\\
v & E & v & 0 & 0 & 0 & 0 & 0 & 0 & 0 & 0\\
0 & v & 2E & v & 0 & 0 & 0 & 0 & 0 & 0 & 0\\
0 & 0 & v & 3E & v & 0 & 0 & 0 & 0 & 0 & 0\\
0 & 0 & 0 & v & 4E & v & 0 & 0 & 0 & 0 & 0\\
0 & 0 & 0 & 0 & v & 5E & v & 0 & 0 & 0 & 0\\
0 & 0 & 0 & 0 & 0 & v & 6E & v & 0 & 0 & 0\\
0 & 0 & 0 & 0 & 0 & 0 & v & 7E & v & 0 & 0\\
0 & 0 & 0 & 0 & 0 & 0 & 0 & v & 8E & v & 0\\
0 & 0 & 0 & 0 & 0 & 0 & 0 & 0 & v & 9E & v\\
0 & 0 & 0 & 0 & 0 & 0 & 0 & 0 & 0 & v & 10E
\end{bmatrix}
\end{gather*}
We introduced a transition operator T which was defined by its matrix
elements \textless{}n T (n+1)\textgreater{} =1. We showed results
for v= 0.5, 1, 2, and 3. In all cases the transition strength as a function
of excitation energy displayed to an excellent degree an exponential
decrease. That is to say on a log plot we got a good fit to a straight
line with a negative slope for all 4 v's considered.

In this work we consider many more values of v both very small and
very large. Will the exponential behavior persist? We will also consider
a different transition operator \textless{}n T(n+1)\textgreater{}
= $\sqrt{(n+1)}$. One gets this if one uses harmonic oscillator
wave functions and the dipole operator x. Which transition operator
is more realistic is complicated. In the standard vibrational model
of nuclei $\sqrt{n}$ coupling appears in most theories. However it
was noted by S.J.Q. Robinson et al.{[}4{]} that even when one gets
equally spaced levels in a shell model calculation e.g. $^{92}$Pd,
the B(E2)'s from n to (n-1) do not rise as n beyond J=4. The ``constant''
scenario is close to the truth.

It should be noted that although we use a $\sqrt{(n+1)}$ transition element
suggested by a dipole operator with harmonic oscillator wave functions, the
matrix model is not really an oscillator model. In the latter there is no
coupling, v, between neighboring states since they have opposite parities.
It is best to regard this as a model in its own right.

\section{The calculation}

We choose E of table 1 to be 1 MeV. The only relevant parameter is v/E.
We show figures for the two transition operators above. We choose ``decimal
values'' of v, namely v=0.01, 0.1,1,10,100,1000,10000,100000, and
1000000. For the case \textless{}n T(n+1)\textgreater{} = constant, we
do not include the last n (=10) in the fit because, as mentioned in
ref {[}1{]} the transition matrix element to this state from ground
vanishes, and so on a log plot we get minus infinity. However, for
the case \textless{n T(n+1)\textgreater{}} = $\sqrt{n}$ we include
the last point at least for small v.

The first nine figures correspond to the constant transition and the
next nine to the dipole transition operators.

\section{Results for \textless{}n T (n+1)\textgreater{} = constant (=1).}

The eigenfunctions of the matrix shown in section 1 are shown as row vectors
($a_{1}$, $a_{2}$, $a_{3}$, $a_{4}$, $a_{5}$, $a_{6}$, $a_{7}$, $a_{8}$, $a_{9}$, $a_{10}$, $a_{11}$). Here $a_{i}$ is the probability amplitude that the
system is in the basis state | i >. The $a_{i}$ are normalized such that sum $a_{i}^{2}$=1.

In this section we consider the case where the transition from one basis state is non
vanishing only to it's nearest neighbors, all other transition matrix elements vanish.
Here in sec 3 we have
\textless{}nT(n+1)\textgreater{} = constant which we take to be one.
The expression for the transition matrix element is:
$$
((a_{1}b_{2}+,..+a_{10}b_{11}) + (b_{1}a_{2}+....+.b_{11}a_{10}))*\textless{}n T (n+1)\textgreater{}
$$
with the last factor taken to be unity. (Eq. 1) 

For v=0.01 we make a linear fit but exclude the first, eighth, and ninth points. We note
that these three points lie above the line formed by the other 6. For v = 0.1,1 and 10 we show linear fits for the
first 9 points. We see at v=10 that irregularities begin to appear.
We have the beginning of a phase transition which manifests itself as 
an even-odd effect as a function of n. For v=100,
1,000, 10,000, 100,000, and 1,000,000 we get 2 nearly parallel lines on a 
log plot (i.e. 2 different exponential drops) one for even n and one for 
odd n. We make a linear fit to both, excluding the first point. The results are shown in Figure 1. 
As was pointed out in ref [1] the transition rate to the 
10th state (last one) is zero and therefore on a log plot would be at 
minus infinity. It was therefore not included in the fit.

\section{Results for \textless{}n T (n+1)\textgreater{} = $\sqrt{(n+1)}$}

We take the expression (Eq. 1) and multiply each term by $\sqrt{(n+1)}$,
where n+1 is the lower of the two indices of a and b. For small v the strength
is reasonably exponential for all points. This is in contrast to the constant
transition where the last value on the log plot was minus infinity. However
for very large v, e.g. v=1000, the last point drops down well below the
exponential fit for the first 9 points. Here one should not include the last
point in a linear fit. For large v there are some irregularities relative to a
linear fit. However, we do not get the striking behavior that was obtained
for a constant transition element, as discussed in Sec 3. That is to say we
do not see two nearly parallel lines. Results are shown in Figs 2.

Note that for very large coupling, i.e. large v, the pattern does not change,
the figures all look the same. However, the energy scale (horizontal axis) is
changing.

\begin{figure}[h]
\caption{Log (base e) of strength for \textless{}n T (n+1)\textgreater{}=1 with linear least squares fit for v = 0.01 to v=1000000}
\includegraphics[scale=0.31]{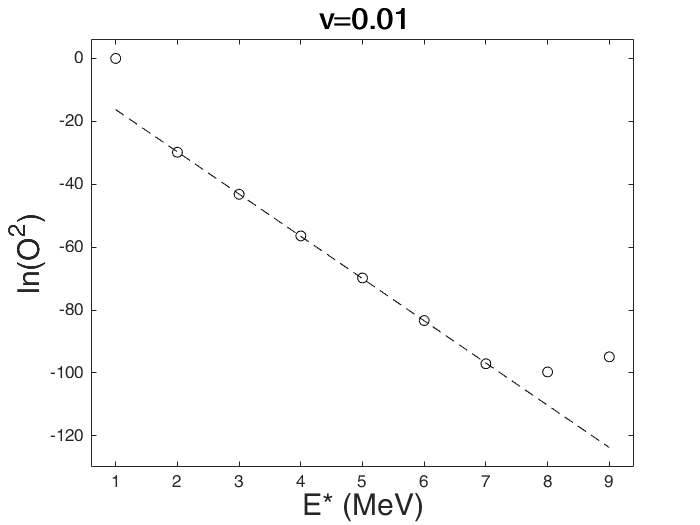}
\includegraphics[scale=0.31]{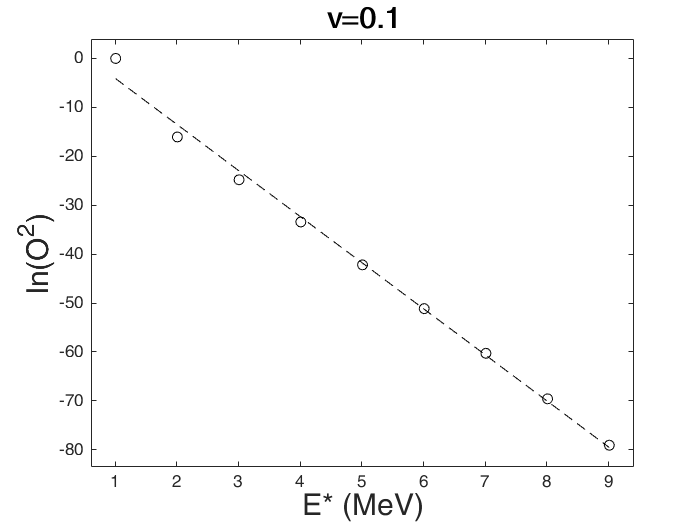}
\includegraphics[scale=0.31]{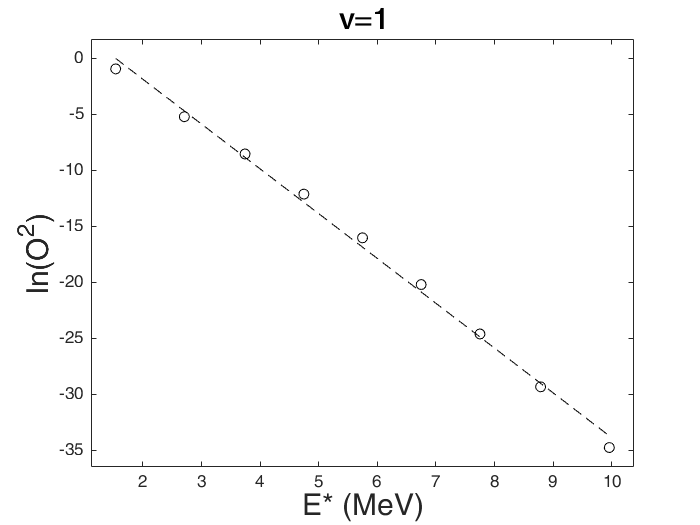}

\includegraphics[scale=0.31]{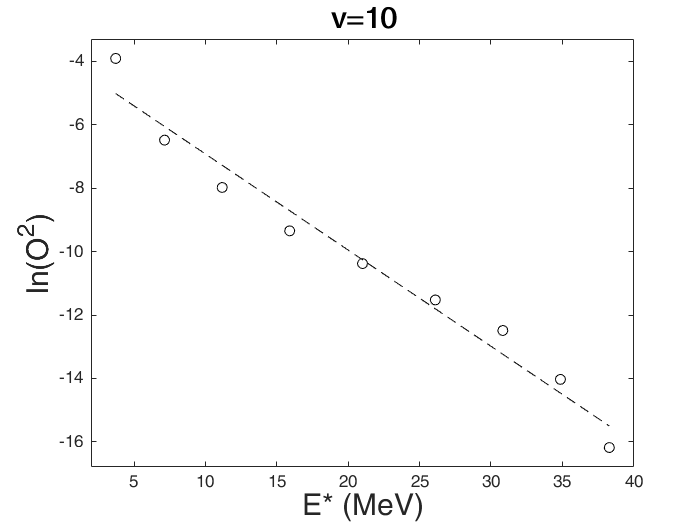}
\includegraphics[scale=0.31]{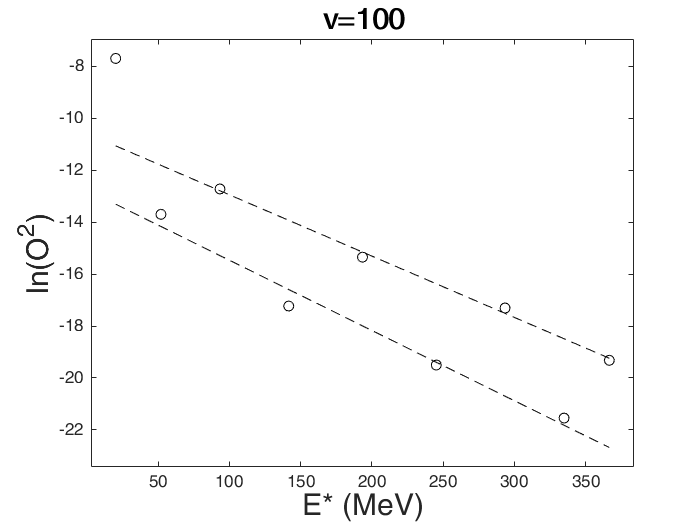}
\includegraphics[scale=0.31]{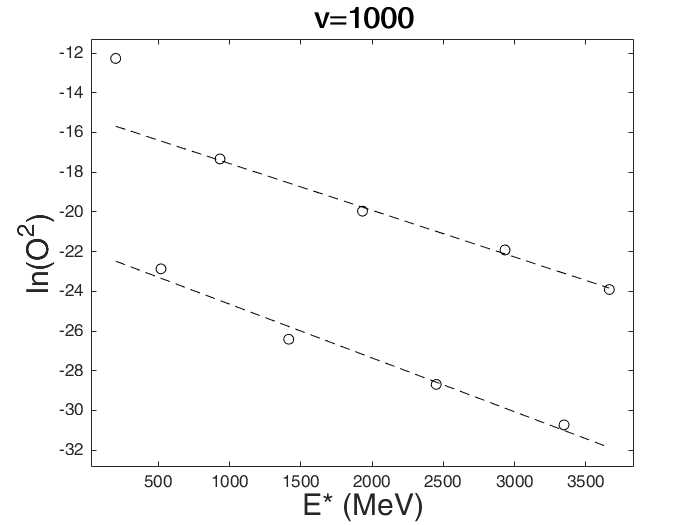}

\includegraphics[scale=0.31]{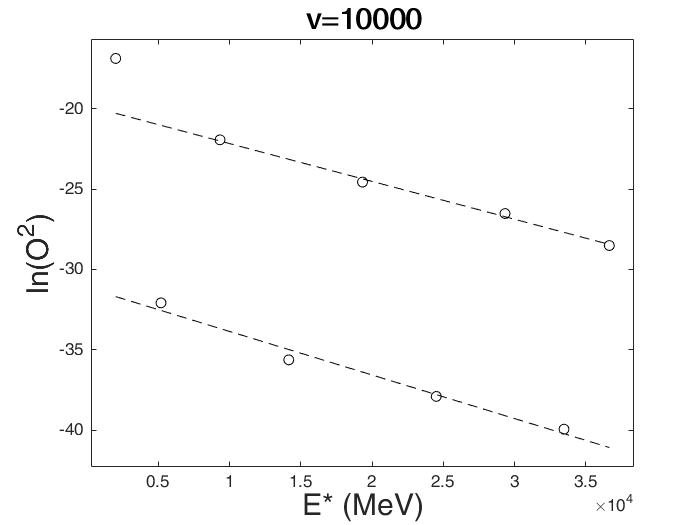}
\includegraphics[scale=0.31]{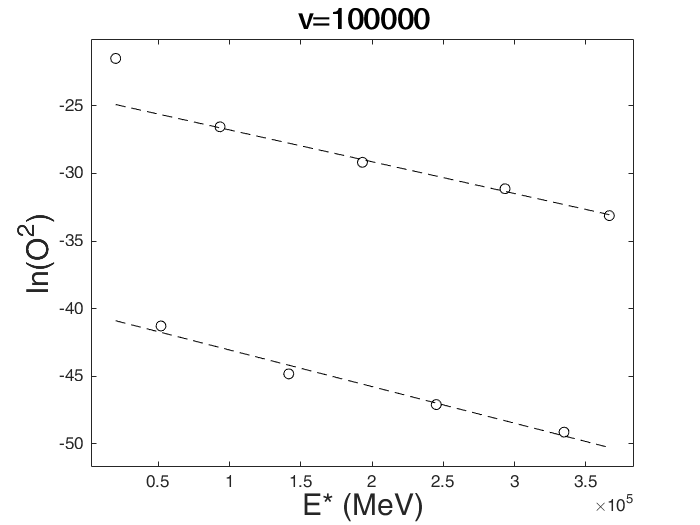}
\includegraphics[scale=0.31]{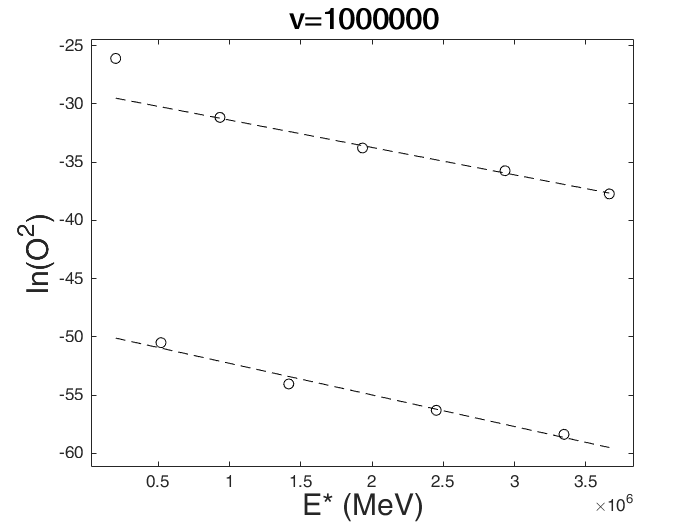}
\end{figure}

\begin{figure}[h]
\caption{Log (base e) of strength for \textless{}n T (n+1)\textgreater{}=$\sqrt{(n+1)}$ with linear least squares fit for v = 0.01 to v=1000000}
\includegraphics[scale=0.31]{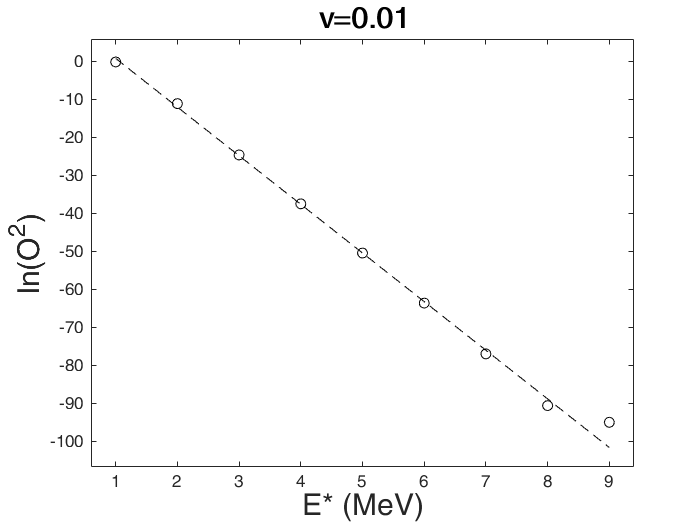}
\includegraphics[scale=0.31]{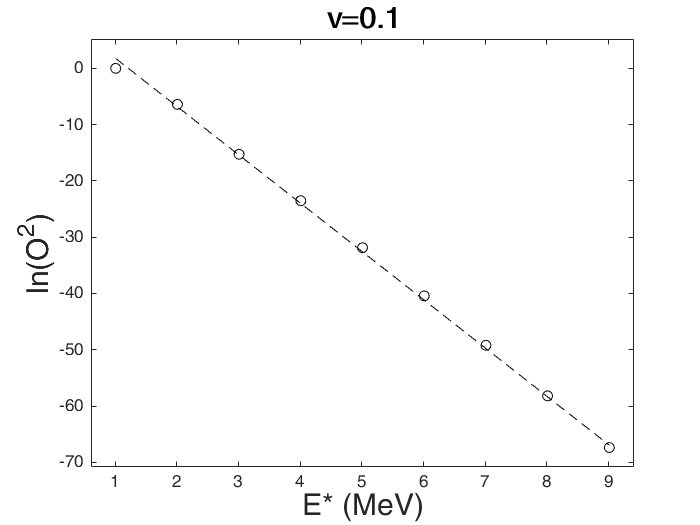}
\includegraphics[scale=0.31]{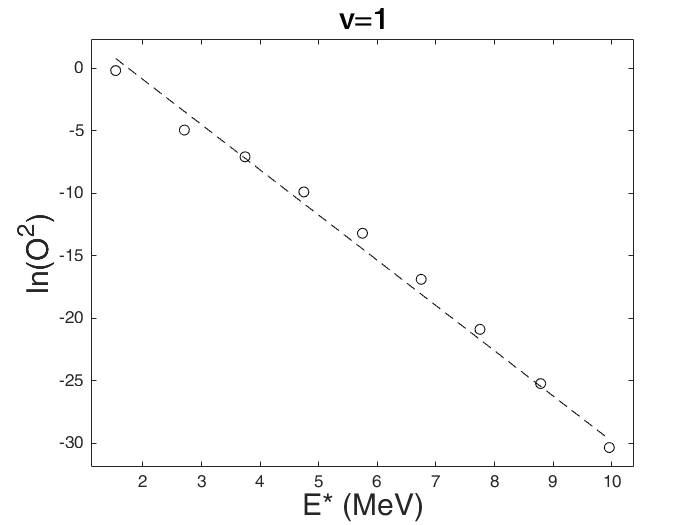}

\vspace{0.5cm}

\includegraphics[scale=0.31]{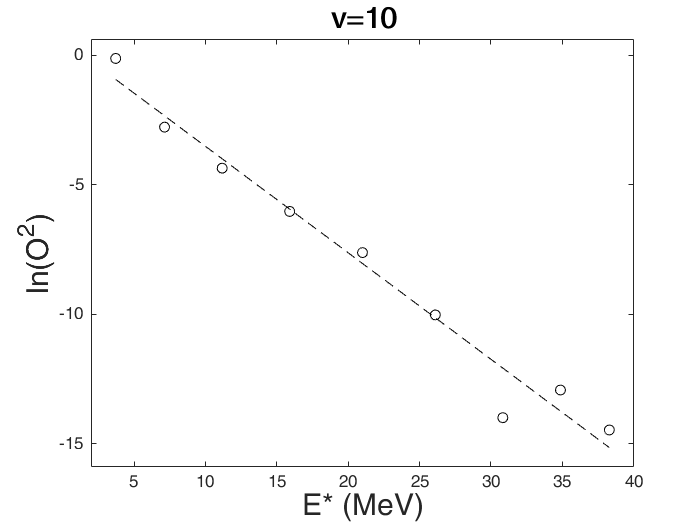}
\includegraphics[scale=0.31]{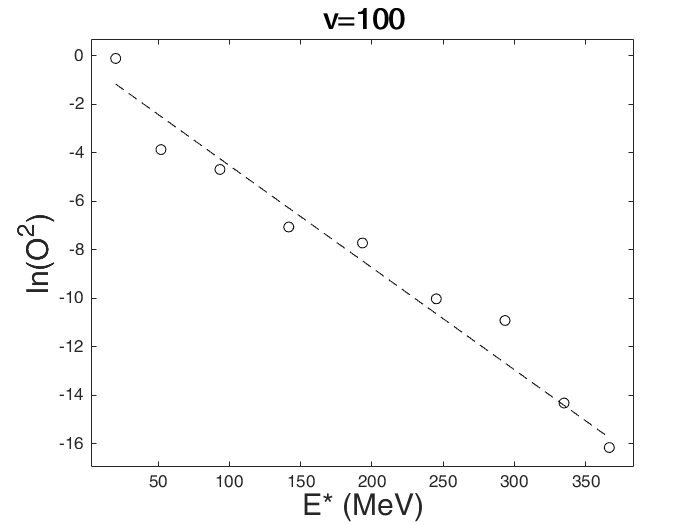}
\includegraphics[scale=0.31]{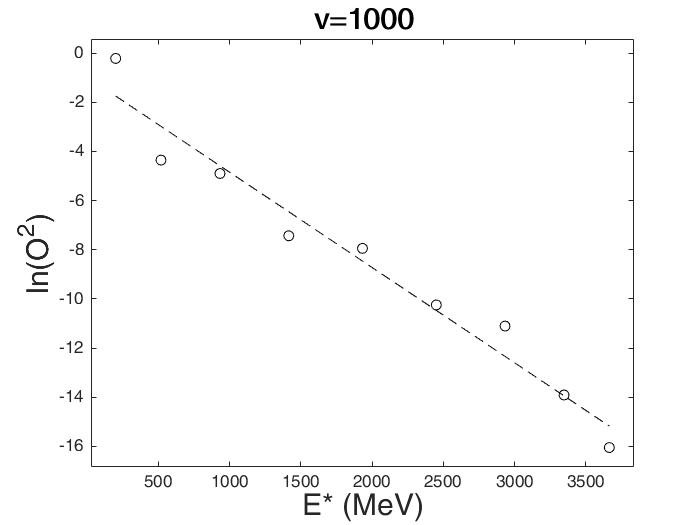}

\vspace{0.5cm}

\includegraphics[scale=0.31]{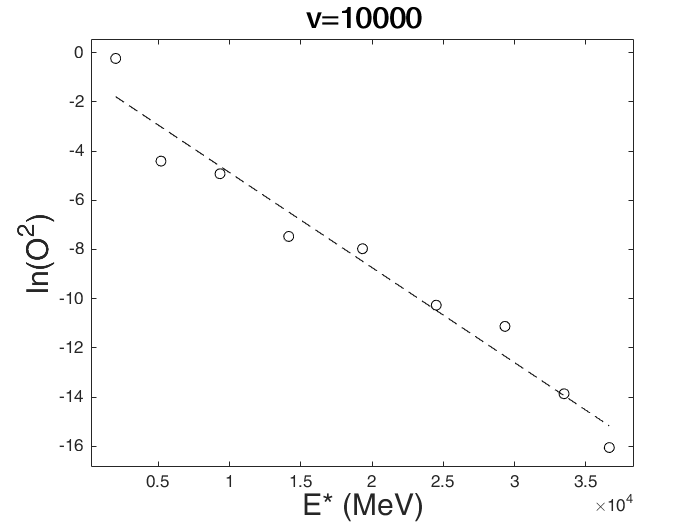}
\includegraphics[scale=0.31]{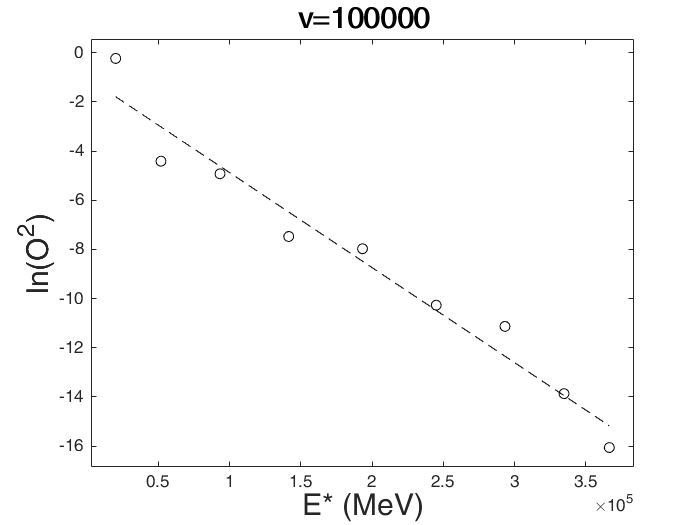}
\includegraphics[scale=0.31]{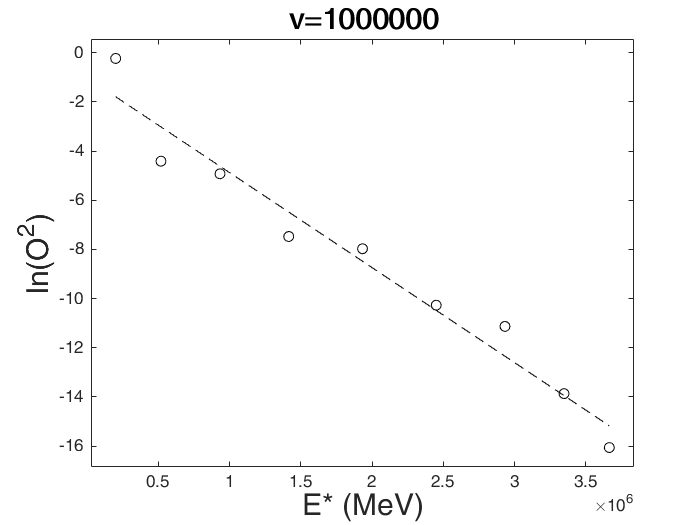}
\end{figure}

\FloatBarrier

\section{Weak and Strong Coupling Limits for the case \textless{}n T (n+1)\textgreater{} = constant}
In the case where there is no coupling, i.e. v=0, the transition strength for the case
\textless{}n T (n+1)\textgreater{} = 1 from the ground state would all go to the first excited state and the value would be O=1. However if we make v very small but not 0, e.g. v=0.01, we find some interesting but approximate observations. First, the ground state wave function amplitudes, which we denote as $a_{i}$, have the following approximate structure:
$$
a_{1}=1,  a_{2}=-v,  a_{3}=v^{2}/2!, a_{4}=-v^{3}/3!, ... , a_{n}=(-v)^{n-1}/(n-1)!
$$
Second, the excited state wave function components, denoted $b_{i}$, involve combinations of the ground state wave function components, $a_{n}$. The first four excited state wave functions are shown in Table I.

\begin{table}[h]
\caption{Wave Functions of Excited States (1 denotes first excited state}
\centering
\begin{tabular}{cccccccccccc}
\hline
State & $b_{1}$ & $b_{2}$ & $b_{3}$ & $b_{4}$ & $b_{5}$ & $b_{6}$ & $b_{7}$ & $b_{8}$ & $b_{9}$ & $b_{10}$ & $b_{11}$ \\
\hline
\hline
1 & -$a_{2}$ & $a_{1}$ & $a_{2}$ & $a_{3}$ & $a_{4}$ & $a_{5}$ & $a_{6}$ & $a_{7}$ & $a_{8}$ & $a_{9}$ & $a_{10}$ \\
2 & $a_{3}$ & -$a_{2}$ & $a_{1}$ & $a_{2}$ & $a_{3}$ & $a_{4}$ & $a_{5}$ & $a_{6}$ & $a_{7}$ & $a_{8}$ & $a_{9}$ \\
3 & -$a_{4}$ & $a_{3}$ & -$a_{2}$ & $a_{1}$ & $a_{2}$ & $a_{3}$ & $a_{4}$ & $a_{5}$ & $a_{6}$ & $a_{7}$ & $a_{8}$ \\
4 & $a_{5}$ & -$a_{4}$ & $a_{3}$ & -$a_{2}$ & $a_{1}$ & $a_{2}$ & $a_{3}$ & $a_{4}$ & $a_{5}$ & $a_{6}$ & $a_{7}$ \\
\hline
\end{tabular}
\end{table}

The expression for the transition amplitude O is:
$$
O =  (a_{1}b_{2} + a_{2}b_{3} + ... + a_{10}b_{11}) + (a_{2}b_{1} + a_{3}b_{2} + ... + a_{11}b_{10})
$$
Because the excited state amplitudes are approximately related to the ground state amplitudes we get many cancellations in evaluating O. Here we show the approximate values (to a first degree) of the transition strength of the ground state to the first four excited states.

First excited state:
$$
O \approx a_{1}b_{2} = a_{1}a_{1} = 1
$$
$$
ln(O^{2})=0 \quad \textrm{[EXACT -0.0002]}
$$

Second excited state:
$$
O \approx a_{3}b_{4} + b_{3}a_{4} = a_{3}a_{2} + a_{1}a_{4}
$$
$$
ln(O^{2})=-28.4419 \quad \textrm{[EXACT -29.8287]}
$$

Third excited state:
$$
O \approx a_{4}b_{5} - b_{4}a_{5} = a_{4}a_{2} - a_{1}a_{5}
$$
$$
ln(O^{2})=-41.0002 \quad \textrm{[EXACT -43.1978]}
$$

Fourth excited state:
$$
O \approx a_{5}b_{6} + b_{5}a_{6} = a_{5}a_{2} + a_{1}a_{6}
$$
$$
ln(O^{2})=-52.0431 \quad \textrm{[EXACT -56.4379]}
$$

The pattern to the nth excited state (excluding the first) is:
$$
O = a_{n+1}a_{2} + (-1)^{n}a_{1}a_{n+2} = v^{n+1}[(-1)^{(n+1)}/n! - 1/(n+1)!]
$$

Although the approximate fit is not great it has all the correct features of the exact calculation. These results were not rigorously derived. Rather they were obtained by a careful observations of the wave function output.

We next consider the strong coupling limit which has already been discussed in ref[1]
but is here inserted for completeness. We can reach this limit either by making v very
large or by setting E to zero. As noted in ref[1] when <n T(n+1)> = constant, all
transitions vanish in this strong coupling limit. This was explained by the fact that in this
limit the transition matrix is proportional to the Hamiltonian matrix of sec [1].
This proportionality is not present for the case \textless{}n T (n+1)\textgreater{} = $\sqrt{(n+1)}$ and indeed all
transitions do not vanish for this latter case.

\section{Addtional Comments}

There are other matrix models which address problems related to but
different from what we have here considered. As previously mentioned in ref {[}1{]},
Bohr and Mottelson {[}5{]} use matrix models to derive the Breit-Wigner
formula for a resonance. Brown and Bolsterli described the giant dipole
resonances in nuclei in a schematic model using a delta interaction {[}6{]}.
In that work they made the approximation that certain radial integrals
were constant. Abbas and Zamick {[}7{]} removed this restriction. Generally
speaking matrix models are very useful for casting insights into the
physics of given problems where the more accurate but involved calculations
fail.

Relative to the first paper {[}1{]} we confirm that decreasing exponential
behavior occurs for a wide range of reasonable parameters for both
types of couplings here considered. However, when we widen the range
of parameters other behaviors occur. One of the most striking examples
is the case v=100 with constant transition operator. In that case
we get two exponentials - one for even n and one for odd n. On a log
plot the two lines are nearly parallel. This behavior also persists
over a wide range e.g. to v=1000.

We close by noting that although this presentation may seem somewhat
mathematical, the idea that strength distributions often display exponential
behavior emanated from physical problems concerning magnetic dipole
excitations, some of the references being {[}8-11{]}. Exponential decreases have been noted and calculated by Schwengner
et al. {[}12{]} in processes that are the inverse of electron excitation,
e.g. photonuclear reactions. Here the reduced M1 decay probabilities
of an excited state are tabulated as a function of the gamma ray energies.
There is an exponential fall off with increasing gamma ray energy. This
differs from our work here in that we have one fixed initial
state (or if we invert things one final state). In the photonuclear
case one considers a cascade of gamma rays from any state to any
other state. Related works by Brown et al. {[}13{]}, Schwengner et al.
{[}14{]}, Siega {[}15{]} and Karampagia et al. {[}16{]} also deal with
these processes. The exponential fall off in these works support our
contention that such behavior is widespread.


\begin{thebibliography}{11}
\bibitem{key-1}L.Zamick and A. Kingan, Int. Jour. Mod. Phys. E,27,1850064 (2018)

\bibitem{key-2}W. Heisenberg, Z. Phys. 33, 879\textendash 893 (1925)

\bibitem{key-3}M. Born and P. Jordan, Z. Phys. 34, 858\textendash 888
(1925)

\bibitem{key-4}S.J.Q. Robinson, T. Hoang, L.Zamick, Y.Y. Sharon,
A. Escuderos Journal-ref: Phys. Rev. C89, 014316 (2014)

\bibitem{key-5} A Bohr and B.R. Mottelson, Nuclear Structre II: Nuclear
Deformation,World Scientific,Singapore (1975).

\bibitem{key-6}G.E. Brown and M. Bolsterli, Phys. Rev. Lett. 3,472
(1959)

\bibitem{key-7} Afsar Abbas and Larry Zamick ,Phys. Rev. C \textbf{22}, 1755 (1980)
( 1980). 

\bibitem{key-8} D. Bohle, A. Richter, W. Steffen, A. E. L. Dieperink,
N. Lo Iudice, F. Palumbo and O. Scholten, Phys. Lett. B 137, 27 (1984)

\bibitem{key-9}K.Heyde, P.von Neumann-Cosel and A. Richter Rev.Mod.Phys.
82, 2365 (2010)

\bibitem{key-10} A. Kingan, M.I. Quinonez, X. Yu and L. Zamick (1984) arXiv:1707.00266

\bibitem{key-11}A. Kingan, and L.Zamick, arXiv:1803.00645

\bibitem{key-12}R. Schwengner, S. Frauendorf and A.C. Larson, Phys. Rev. Lett. 111, 232504 (2013)

\bibitem{key-13}B. Alex Brown and A.C. Larsen Phys. Rev. Lett. 113, 252502 (2014) 

\bibitem{key-14}R. Schwengner,S. Frauendorf and B.A. Brown, Phys.Rev.Lett. 118,092502 (2017)

\bibitem{key-15}K. Siega, Phys. Rev. Lett 119 ,052502 (2017)

\bibitem{key-16}S. Karampagia, B.A. Brown and V. Zelevinsky, Phys. Rev C95, 024322 (2017)

\end{thebibliography}
\end{document}